\documentclass[twocolumn]{aastex631}

\usepackage[caption=false]{subfig}
\usepackage{forloop}
\usepackage{lipsum}

\shorttitle{Detecting anomalous images}
\shortauthors{Alonso et al.}

\graphicspath{{./}{figures/}}

\begin{document}

\title{Detecting anomalous images in astronomical datasets}

\correspondingauthor{Jun Zhang}
\email{betajzhang@sjtu.edu.cn}

\author{Pedro Alonso}
\affiliation{Department of Astronomy, Shanghai Jiao Tong University, Shanghai 200240, China}
\affiliation{Shanghai Key Laboratory for Particle Physics and Cosmology, Shanghai 200240, China}

\author{Jun Zhang}
\affiliation{Department of Astronomy, Shanghai Jiao Tong University, Shanghai 200240, China}
\affiliation{Shanghai Key Laboratory for Particle Physics and Cosmology, Shanghai 200240, China}

\author{Xiao-Dong Li}
\affiliation{School of Physics and Astronomy, Sun Yat-Sen University, Guangzhou 510297, China}
\affiliation{CSST Science Center for the Guangdong–Hong Kong–Macau Greater Bay Area, SYSU, Zhuhai 519082, China}

\begin{abstract}
Environmental and instrumental conditions can cause anomalies in astronomical images, which can potentially bias all kinds of measurements if not excluded. Detection of the anomalous images is usually done by human eyes, which is slow and sometimes not accurate. This is an important issue in weak lensing studies, particularly in the era of large scale galaxy surveys, in which image qualities are crucial for the success of galaxy shape measurements. In this work we present two automatic methods for detecting anomalous images in astronomical datasets. The anomalous features can be divided into two types: one is associated with the source images, and the other appears on the background. Our first method, called the Entropy Method, utilizes the randomness of the orientation distribution of the source shapes and the background gradients to quantify the likelihood of an exposure being anomalous. Our second method involves training a neural network (autoencoder) to detect anomalies. We evaluate the effectiveness of the Entropy Method on the CFHTLenS and DECaLS DR3 data. In CFHTLenS, with 1171 exposures, the Entropy Method outperforms human inspection by detecting 12 of the 13 anomalous exposures found during human inspection and uncovering 10 new ones. In DECaLS DR3, with 17112 exposures, the Entropy method detects a significant number of anomalous exposures while keeping a low false positive rate. We find that although the neural network performs relatively well in detecting source anomalies, its current performance is not as good as the Entropy Method.
\end{abstract}


\section{Introduction}

\label{section:Introduction}
Anomalous images can be caused by atmospheric, environmental, or instrumental conditions. They can be categorized into source and background anomalies. Source anomalies are prominent in the vicinity of bright sources, distorting their shapes, whereas background anomalies generally manifest systematic wave-like patterns that distort the entire background of the CCD images. Fig.~\ref{fig:source_anomalies_examples} and fig.~\ref{fig:back_anomalies_examples} show several examples of source and background anomalies, respectively. 

Anomalous images can be a source of systematics in many astronomical measurements. It could be a particularly important issue in weak lensing studies if they are mistakenly included in the dataset.
This is because the cosmic shear signal is only of the order of a few percent, highly sensitive to the quality of the images. In past surveys, the number of exposures was small enough to check each one individually and discard the anomalous ones. More recent surveys, such as SDSS \citep{2000AJ....120.1579Y}, DES \citep{2005astro.ph.10346T}, CFHTLenS \citep{2013MNRAS.432.2433H}, HSC \citep{2018PASJ...70S...4A}, or KiDS \citep{2013ExA....35...25D}, typically produce tens of thousands of exposures, making human inspection impractical. As a result, the development of algorithms that can automatically detect anomalous images is becoming crucial. 

Anomaly detection techniques are typically classified as supervised or unsupervised. Supervised methods use labeled data to directly learn the difference between normal and anomalous data. Common supervised anomaly detection techniques include support vector machines \citep[SVMs:][]{708428} and neural networks. In most real case scenarios, however, the amount of anomalous data is generally much less than normal data, making it difficult to collect enough anomalous data for training. Therefore, anomaly detection algorithms have to rely solely on the information about normal data. Unsupervised learning algorithms first learn the common features that describe normal data and then use an anomaly detector to judge how similar new data is to the normal one. If the features of the new data are very different from those of normal data, it is classified as anomalous. Well-known unsupervised techniques used in anomaly detection include principal component analysis \citep[PCA:][]{article1}, isolation forest \citep{4781136}, autoencoders \citep{Rumelhart1986LearningRB, 2022arXiv220103898M}, k-Nearest Neighbors \citep[KNN:][]{article2}, and K-means \citep{MacQueen1967SomeMF}. In particular, visual anomaly detection is an active field of machine learning (see \cite{2021arXiv210913157Y} for a review) with applications in many fields, including astronomy \citep{2022RAA....22h5006H}.

In this project, we present two methods to detect anomalous images in astronomical datasets. Our first method is analytical, and we call it the Entropy Method. Our second method uses a neural network called autoencoder to detect anomalies in an unsupervised way. We show the performances of the two methods and compare them. We search for anomalies in the CFHTLenS and DECaLS DR3 datasets.

The structure of this paper is as follows. \S\ref{sec:data} describes our sources of data and introduces sources and background anomalies. \S\ref{section:Methods_analytical} presents the Entropy Method and its results. In \S\ref{section:Methods_autoencoder} we introduce our deep learning method and its results, and we conclude in \S\ref{section:Conclusion}.

\section{Data and examples of anomalies}
\label{sec:data}
In this paper, our main focus is on identifying anomalous exposures within two significant astronomical datasets. First, we use data from the The Canada-France-Hawaii Telescope Lensing Survey \citep[CFHTLenS:][]{2012MNRAS.427..146H, 2013MNRAS.433.2545E}, which was specifically design for weak lensing studies as part of the CFHT Legacy Survey. Second, we use data from the third data release of the Dark Energy Camera Legacy Survey (DECaLS DR3). DECaLS is one of the imaging projects of the Dark Energy Spectroscopic Instrument \citep[DESI:][]{2016arXiv161100036D} and it is extensively used in many astronomical studies, including weak lensing.

CFHTLenS includes optical data in 4 fields (W1, W2, W3, W4), covering a total of $154$ $\text{deg}^2$ in 5 optical bands (u, g, r, i, z), with the i-band showing better seeing conditions than the rest. We search for anomalies in the i-band of CFHTLenS, which consists of 1171 exposures. Each exposure is composed of 36 CCD images.

DECaLS DR3 includes optical data in three bands (g, r, z), covering a total of $4300$ $\text{deg}^2$ in g-band, $4600$ $\text{deg}^2$ in r-band, and $8100$ $\text{deg}^2$ in z-band. We search for anomalies in the g, r, and z bands, which are composed of 3919, 5712, and 7481 exposures, respectively. Each exposure consists of 61 CCD images.

The i-band of CFHTLenS has undergone visual inspection, enabling us to directly benchmark the performance of our methods against human inspection. Furthermore, DECaLS DR3 provides us with a huge number of exposures to further test the performance of our methods. Combining the datasets from CFHTLenS and DECaLS DR3 forms an extensive and comprehensive collection of data, enhancing the robustness of our analysis and enabling us to draw more meaningful conclusions about anomalous exposures and weak lensing studies.

In the following we describe source and background anomalies and provide some examples from CFHTLenS and DECaLS DR3.

\subsection{Source anomalies}
\label{sec:source_anomalies_intro}
\begin{figure}
    \begin{center}
    \includegraphics[width=0.49\textwidth]{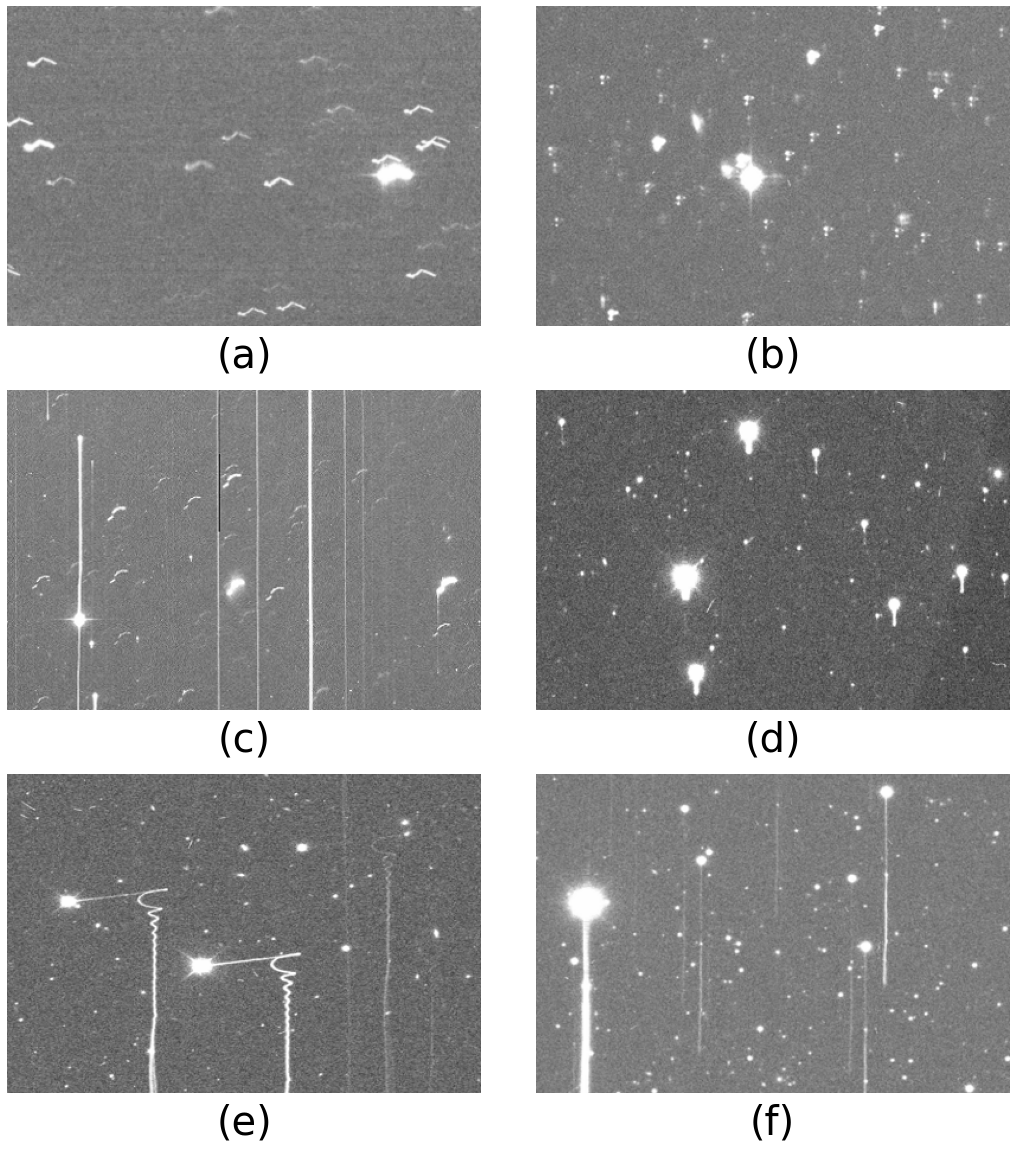}
	\caption{Examples of source anomalies in CFHTLenS and DECaLS DR3.}
	\label{fig:source_anomalies_examples}
    \end{center}
\end{figure}

Fig.~\ref{fig:source_anomalies_examples} shows examples of source anomalies, which are defined as those anomalies that distort the sources or their vicinity across entire exposures. They can be caused by various optical aberrations, such as coma, which produces sources with comet-like shapes, or astigmatism, which causes elongations on the sources. Other examples of optical aberrations that can distort the shape of sources include spherical aberration and field curvature. Thankfully, these optical aberrations have been extensively studied and modern telescopes can correct for them. Nevertheless, it is possible that most source anomalies in our data may be attributed to small shifts on the focal plane of the telescope during observations.
Depending on the magnitude of this shift, it can introduce elongations on the sources (panels a and c of fig.~\ref{fig:source_anomalies_examples}) or even generate multiple source images (panel b of fig.~\ref{fig:source_anomalies_examples}). Other types of source anomalies are likely related to problems during the readout process. Vertical bright trails like those found in panels d and f of fig.~\ref{fig:source_anomalies_examples} can be generated when the readout process starts while the shutter is still open. There are other source anomalies with very peculiar forms such as panel e of fig.~\ref{fig:source_anomalies_examples}.

Despite the diverse nature of source anomalies, we have observed that all of them exhibit a certain coherent anisotropy across the entire exposure. This anisotropy distorts the sources or their vicinity along a particular direction. This observation prompts us to utilize the orientation of the sources across the entire exposure as our indicator for detecting source anomalies.

We perform a statistical study of the orientation of all the sources within each exposure to determine the likelihood of that exposure presenting a source anomaly. The detailed description of this methodology is given in \S\ref{section:source_anomalies_method}.

\subsection{Background anomalies}
\label{sec:background_anomalies_intro}
\begin{figure}
    \begin{center}
    \includegraphics[width=0.45\textwidth]{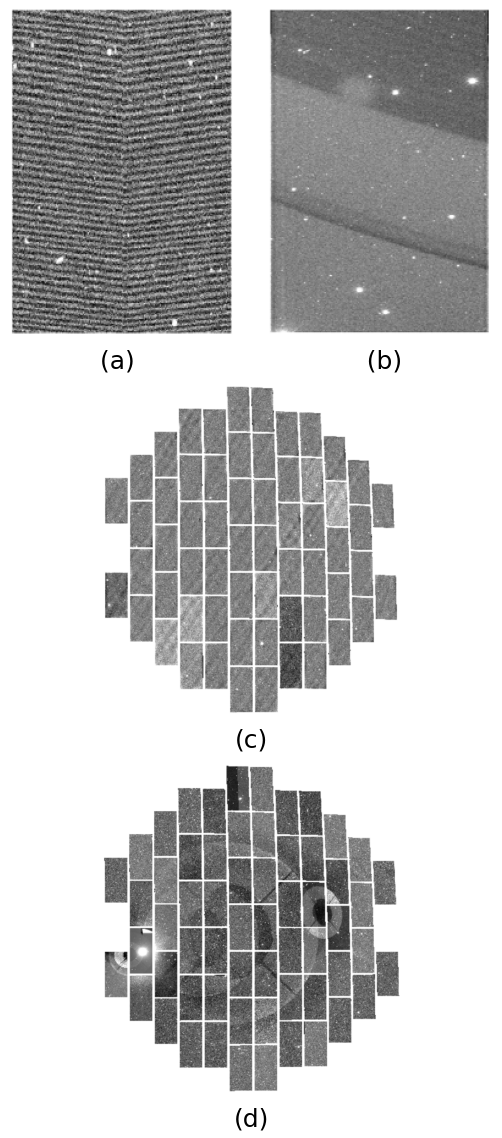}
	\caption{Examples of background anomalies in DECaLS DR3.}
	\label{fig:back_anomalies_examples}
    \end{center}
\end{figure}

Background anomalies can introduce distortions in the CCD images, and there are two main types. First, light reflections inside the telescope can cause the background to present a series of circular shapes that cover a significant part of the exposure. Sometimes this effect is visible on single CCD images (panel b of fig.~\ref{fig:back_anomalies_examples}), and, in other cases, the anomaly only become noticeable when looking at the entire exposure (panel d of fig.~\ref{fig:back_anomalies_examples}). The second type of background anomaly manifests as frequent stripes in the CCDs, likely caused by electronic noise. These stripes exhibit a preferred direction, visible either on single CCDs (panel a of fig.~\ref{fig:back_anomalies_examples}) or when looking at the entire exposure (panel c of fig.~\ref{fig:back_anomalies_examples}).

Similar to source anomaly identification, we utilize the level of statistical anisotropy of background gradients to detect background anomalies. A more detailed explanation of the method is given in \S\ref{section:back_anomalies_method}.

\section{Entropy Method for detecting anomalies} 
\label{section:Methods_analytical}
In \S\ref{sec:data} we have shown that both source and background anomalies distort the CCD images. In fact, these distortions can be quantified in terms of entropy. In a normal image, sources are randomly oriented, which corresponds to a state of high entropy. If the image presents a source anomaly, the majority of sources are oriented towards a particular direction. This corresponds to a state of reduced entropy. In the same way, the background gradients of a normal image should be random. Background anomalies induce spatially coherent distortions, which can significantly reduce the entropy of the background distribution. In this section we present a method that quantifies the entropy of the CCD images and use it as an indicator of the presence of anomalies.

\subsection{Source anomalies}
\label{section:source_anomalies_method}
To quantify the entropy of the sources in terms of their orientation. We define the source orientation as the angle between its long axis and the x-axis, and we call it $\theta$.
For each exposure, we construct a distribution of $\theta$, including all sources in that exposure. In a normal image, the orientation of the sources should be random, resulting in a flat $\theta$ distribution and a state of highest entropy. In the presence of a source anomaly, however, the sources will show a preferred orientation and the $\theta$ distribution will not be flat, corresponding to a state of reduced entropy.

We quantify the anomaly likelihood as the deviation of each $\theta$ distribution from the mean $\theta$ distribution of all exposures, with a higher deviation indicating a higher likelihood of that exposure presenting a source anomaly. We calculate this deviation using the Kullback-Leibler (KL) divergence \citep{Kullback1959InformationTA, Cover1991ElementsOI}, which is commonly used in machine learning and information theory as a measure of the distance between two probability distributions.
For each exposure, we divide its $\theta$ distribution into N bins of equal size and define its KL divergence as,
\begin{equation}
    \label{eq:KL_div_sources}
    \mathrm{KL}(P_i|\overline{P}) = \sum_{j=1}^{N} {P_i}(\theta_j)\cdot \mathrm{ln} \frac{P_i(\theta_j)}{\overline{P}(\theta_j)},
\end{equation}
where $P_i(\theta_j)$ represents the value of the $j^{th}$ bin in the $\theta$ distribution of the $i^{th}$ exposure and $\overline{P}(\theta_j)$ represents the $j^{th}$ bin in the mean $\theta$ distribution of all exposures.

Since it is assumed that the number of anomalous exposures is very small compared to the size of the entire dataset, the mean $\theta$ distribution of all exposures can be safely used as a representation of the distribution of normal exposures.

To build the $\theta$ distribution of each exposure, we follow these steps:
\begin{enumerate}
    \item We center each source on a 64x64 stamp and calculate its ellipticity components as,
    \begin{equation}
        e_1 = \frac{Q_{20}-Q_{02}}{Q_{20}+Q_{02}}
    \end{equation}
    and
    \begin{equation}
        e_2 = \frac{2Q_{11}}{Q_{20}+Q_{02}}.
    \end{equation}
    $Q_{ij}$ are the quadrupole moments defined as,
    \begin{equation}
        Q_{ij} = \sum_{I(r)>I_0} I(r)x^i y^j,
    \end{equation}
    where $r=\sqrt{x^2+y^2}$. The threshold $I_0$ is set as $I_0=0.02I(\vec{r}=0)$.
    \item We calculate the angle between the long axis of the ellipse and the x axis as:
    \begin{equation}
        \theta = \frac{\mathrm{arctan}(e2/e1)}{2}
    \end{equation}
    \item For each exposure, we create a distribution $\theta$ including all the sources in that exposure.
\end{enumerate}

To identify sources, we realize that there are many possible algorithms for background removal and source identification. However, these algorithms are usually computationally expensive. For the purpose of detecting anomalies we do not need a very accurate source identification method, since the anomalies are more noticeable around the brightest sources. Therefore, we do not need to select really faint sources, which require high level source identification models. We aim for a quick and relatively accurate background removal and source identification methods. In addition, it is important to note that, in contrast with common source identification algorithms, we do not require sources to be isolated within their stamp, meaning that other sources or bright pixels can contaminate the stamp. This is because many anomalies present multiple source images (panel b of fig.~\ref{fig:source_anomalies_examples}) or other bright pixels (panels d, e, and f of fig.~\ref{fig:source_anomalies_examples}), and this requirement would directly discard those sources and complicate the anomaly detection.

To build our background model, we select 1000 random points on the CCD image, and discard the brightest 10\% and the darkest 10\% to avoid source pixels and defective areas. We fit the remaining points to a polynomial of second order. This builds our background model that is subtracted from the CCD image.

After the background has been removed, we define each source center as a pixel that meets the following criteria:
\begin{itemize}
    \item Its value is above 4$\sigma$\footnote{We have explicitly tested setting higher thresholds for source selection, such as $10\sigma$ and $30\sigma$, which include only the brightest sources. They lead to very similar results in the detection of anomalies.}.
    \item It is brighter than all the other pixels within a $64$x$64$ stamp centered in it.
    \item The number of pixels above 2$\sigma$ within that stamp is higher than 6.
\end{itemize}

\subsection{Background anomalies}
\label{section:back_anomalies_method}
Ideally, the background of a CCD image should exhibit a uniform distribution, with the background gradients being random, corresponding to a state of highest entropy. Background anomalies, however, introduce coherent features in the background, such as stripes or wave-like patterns, which distort the natural pixel uniformity and reduce the entropy of the background distribution.

Similar to source anomaly detection, we search for a parameter that can quantify the change of entropy in the background. We choose this parameter to be the angle between the gradient at a given background point and the x-axis, and we call it $\theta$. In a normal image, if we sample over a large number of background positions, the $\theta$ distribution should be flat, since the gradients should be random. However, in an image presenting a background anomaly, the gradients exhibit certain preferred orientations, implying a reduced entropy in the $\theta$ distribution. 

The $\theta$ distribution is built following these steps:

\begin{enumerate}
    \item We smooth the background by applying a 10x10 mean kernel on the CCD image, to enhance the wave-like features that define the anomaly and facilitate their detection.
    \item We select 10000 random points from the smoothed image. For each point at position (i, j), we calculate its gradient components as
    \begin{equation}   D_x=\frac{1}{2}\sum_{m=i+1}^{i+5}\sum_{n=j-5}^{j+5} \frac{(I_{m,n}-I_{2i-m, 2j-n})}{m-i}
    \end{equation}
    \begin{equation}
    D_y=\frac{1}{2}\sum_{m=i-5}^{i+5}\sum_{n=j+1}^{j+5} \frac{(I_{m,n}-I_{2i-m, 2j-n})}{n-j},
    \end{equation}
    where $I_{m,n}$ is the value of the pixel at position (m, n).
    From the gradient components we derive the angle $\theta$ between the total gradient and the x-axis as,
    \begin{equation}
        \theta = \mathrm{arctan}\left( \frac{D_y}{D_x} \right)
    \end{equation}
    \item Similar to source anomalies detection, we form a distribution of $\theta$, but this time on a CCD level rather than exposure level. This is because, unlike in source anomalies, we do not expect the same preferred direction for $\theta$ on every CCD of the exposure.
\end{enumerate}

We calculate the KL divergence (eq.~\ref{eq:KL_div_sources}) between the $\theta$ distribution of each CCD and the mean $\theta$ distribution of all CCDs in the dataset. For each exposure, we average the KL divergence of all its CCDs and use this value as our background anomaly likelihood indicator.

\subsection{Results} \label{section:Results_entropy}

\subsubsection{CFHTLenS}
\label{section:Results_entropy_cfhtlens}

\begin{figure}
    \begin{center}
	\includegraphics[width=0.45\textwidth]{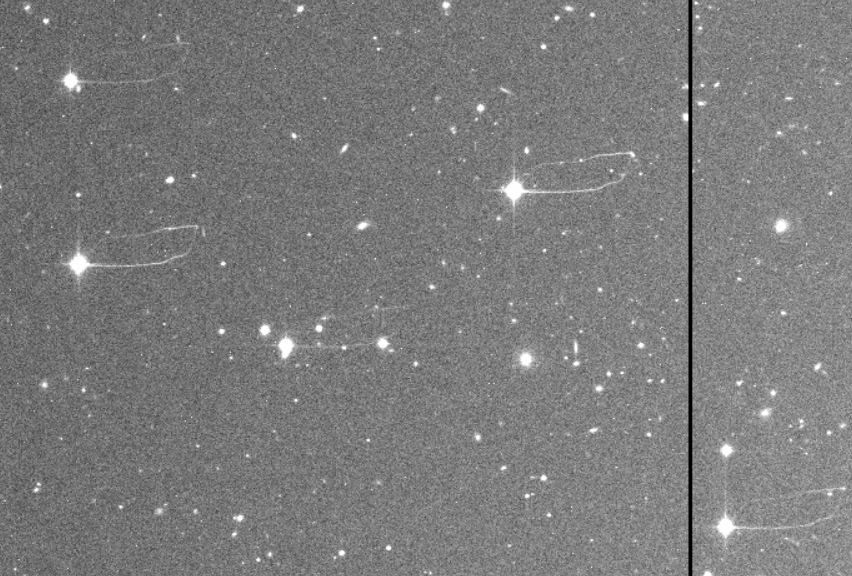}
	\caption{CCD patch of the exposure '792436' of the w3m1m2 field of CFHTLenS.}
	\label{fig:792436}
    \end{center}
\end{figure}

\begin{figure}
    \begin{center}	
    \includegraphics[width=0.45\textwidth]{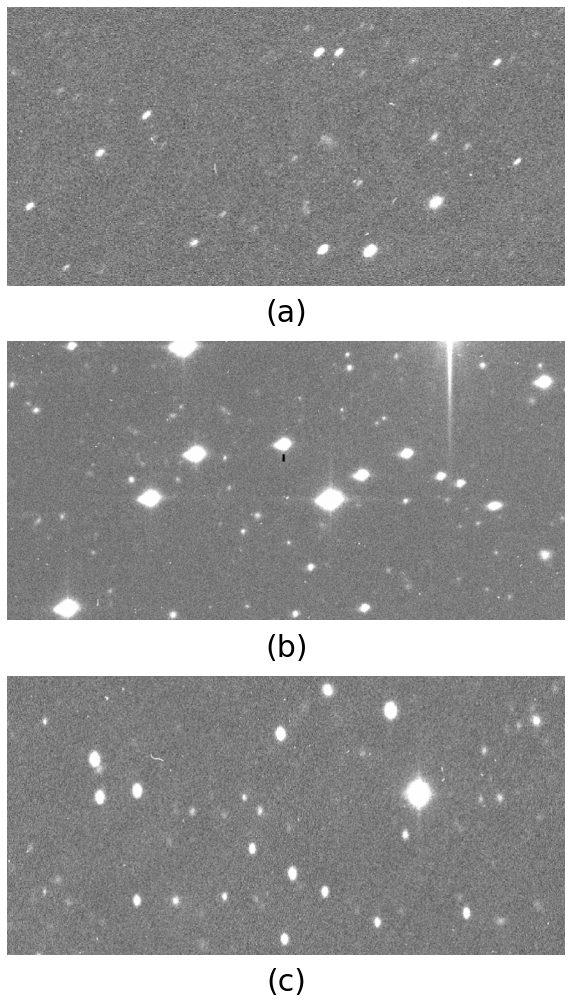}
    \caption{Examples of anomalous exposures in CFHTLenS that were detected by the Entropy Method and missed by human inspection.}
    \label{fig:new_anomalies_cfhtlens}
    \end{center}
\end{figure}

\cite{2019ApJ...875...48Z} reported 13 anomalous exposures in the i-band of CFHTLenS\footnote{"827410" of field w1m2m2, "792617" of w3m1m0, "792436" of w3m1m2, "987104" of w3p2m3, "859948" and "859950" of w4m1p0, and all 7 exposures of w2p2p2}. Applying the Entropy Method to this dataset, we can successfully detect 12 of the 13 exposures identified by \cite{2019ApJ...875...48Z}, only failing to detect the exposure "792436" of w3m1m2 (fig.~\ref{fig:792436}), which presents extremely faint anomaly patterns that are only visible around very bright sources. From the anomalies reported by \cite{2019ApJ...875...48Z}, we detect 5 as source anomalies and 7 as background anomalies (all exposures of field w2p2p2). In addition, we detect 10 new source anomalies; "705392", "705393", and "705394" of field w3m0m1, "732383", "732384", and "732385" of field w2p3p2, "797642" and "797643" of w3p1p2, "775160" of w2p1p3, and "853686" of field w3p3m3.
Fig.~\ref{fig:new_anomalies_cfhtlens} shows a few examples of the newly detected source anomalies. These anomalies present small elongations on the sources, creating a systematic anisotropy across the entire exposure, which is detected by our method but could easily be missed by human inspection. It is also possible that these elongations could be understood as being caused by an elongated PSF. However, as we discuss in \S\ref{section:Results_entropy_decals_dr3}, we believe that these are actual anomalies.

\subsubsection{DECaLS DR3}
\label{section:Results_entropy_decals_dr3}
With more than 17000 exposures, DECaLS DR3 forms a significantly larger dataset than CFHTLenS, thus we do not check all the exposures by eye to evaluate the performance of our method. Instead, we only check the $10\%$ of exposures with the highest likelihood of being anomalous according to our results. We analyze each band (g, r, z) independently.

\begin{figure}
    \begin{center}
	\includegraphics[width=0.49\textwidth]{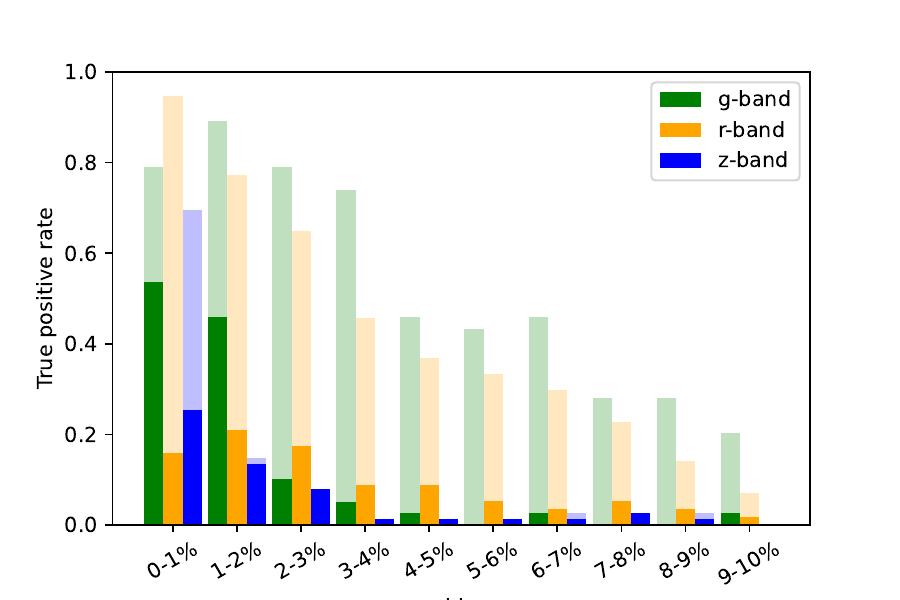}
	\caption{True positive rate for the 10\% of exposures with highest likelihood of presenting source anomalies according to the Entropy Method. The true positive rate is based on human inspection. We divide the exposures into 10 bins of equal size, for each band (g, r, z). Dark colors include only prominent anomalies, whereas light colors also include dubious anomalies.}
	\label{fig:results_sources}
    \end{center}
\end{figure}

Fig.~\ref{fig:results_sources} shows the results for source anomalies. We show the results for the $10\%$ of exposures with highest KL divergence, i.e., highest likelihood of being anomalous. We divide them into 10 bins of equal size and illustrate the true positive rate (TPR) for each bin, based on eye check. As shown in \S\ref{section:Results_entropy_cfhtlens}, eye check can miss many anomalies that can be captured by our method, and therefore here we only use the eye check as a performance indicator, not as the absolute truth.

\begin{figure}
    \begin{center}
	\includegraphics[width=0.45\textwidth]{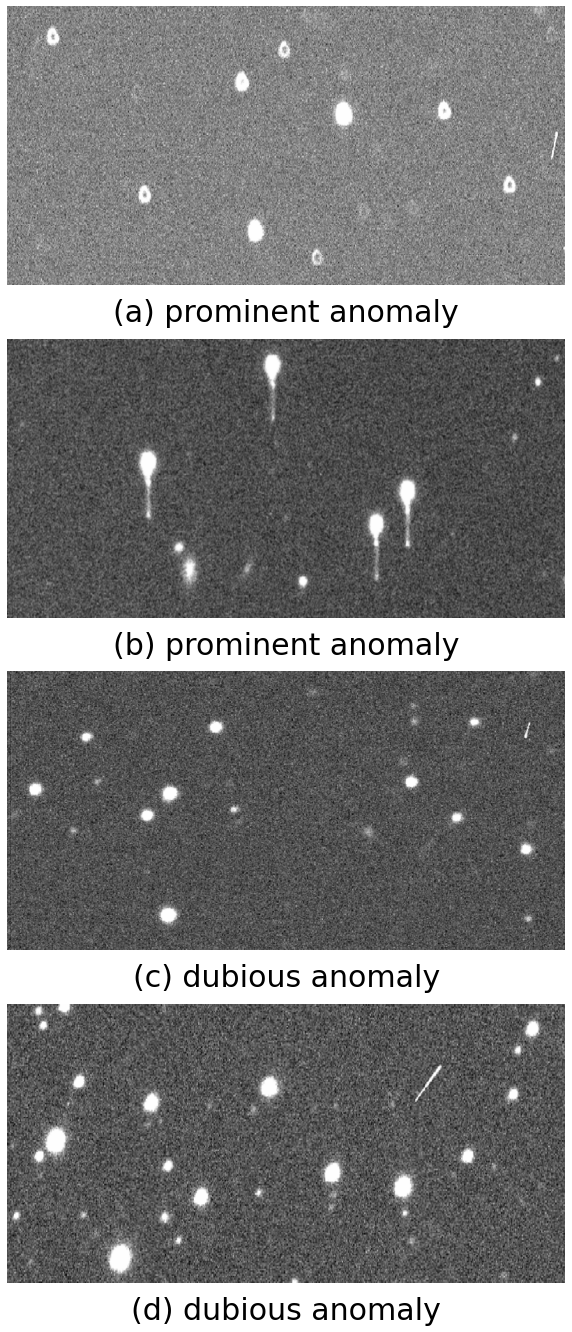}
	\caption{Examples of prominent and dubious anomalous exposures identified in DECaLS by the Entropy Method.}
	\label{fig:anomalies_decals}
    \end{center}
\end{figure}

In fig.~\ref{fig:results_sources}, dark colors represent the TPR including only prominent anomalies, i.e., easily observed by eye and undoubtedly anomalous. Light colors include cases in which the sources present a small but systematic elongation along a particular direction, which is very hard to detect by eye. As mentioned in \S\ref{section:Results_entropy_cfhtlens}, these small elongations could also be seen as caused by an elongated PSF. A strong enough PSF reconstruction technique should be able to handle these elongated sources, but weaker PSF reconstruction techniques might not be able to properly model the PSF and this could introduce bias in shear measurements. We therefore show the TPR for the most restricted case (dark colors) and for the most conservative one (light colors). Fig.~\ref{fig:anomalies_decals} presents four anomalous exposures found in DECaLS DR3, where two of them were classified as prominent anomalies and the other two as dubious.

It is important to note that most of these dubious exposures present very small elongations along the vertical direction. If these anomalies were caused by an elongated PSF, the orientation of the psf should be, although roughly constant along the CCD image, random from exposure to exposure. The preferred vertical direction in most anomalous exposures in the dataset seems to indicate that these elongations are actually caused by anomalies (possibly due to CCD electronics) and not by an elongated PSF.

\begin{figure}
    \begin{center}
	\includegraphics[width=0.49\textwidth]{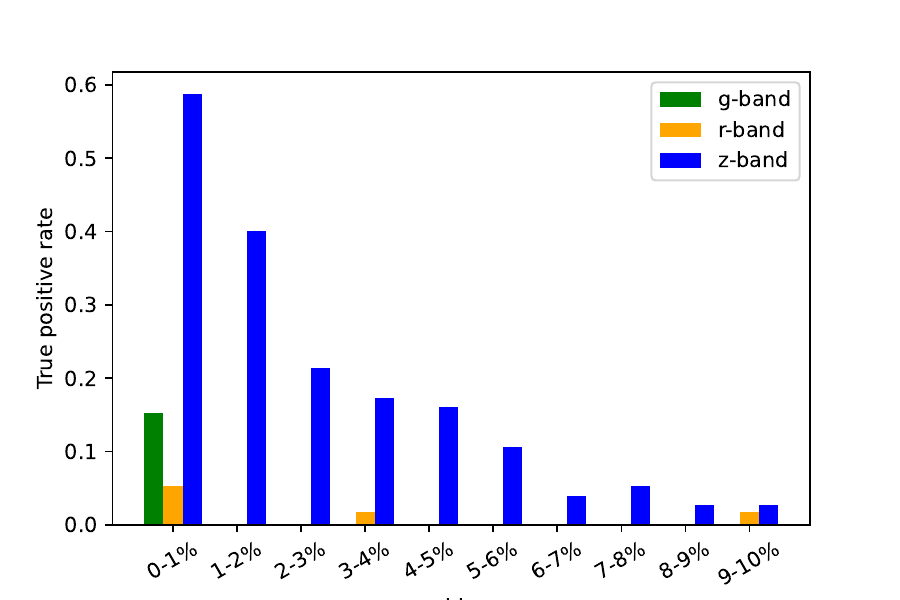}
	\caption{True positive rate for the 10\% of exposures with highest likelihood of presenting background anomalies according to the Entropy Method. The true positive rate is based on human inspection. We divide the exposures into 10 bins of equal size, for each band (g, r, z).}
	\label{fig:results_back}
     \end{center}
\end{figure}

Fig.~\ref{fig:results_back} shows the results for background anomalies. We plot the top $10\%$ of exposures with the highest likelihood of being anomalous, as in source anomalies. However, in this case we do not need to distinguish between obvious and dubious anomalous exposures, since background anomalies are less ambiguous for the human eye.

\section{Deep learning method for detecting anomalies}
\label{section:Methods_autoencoder}
In this method we use a neural network called autoencoder to detect source anomalies. We train the autoencoder to reconstruct source stamps and we use the quality of these reconstructions to quantify the likelihood of exposures presenting source anomalies. In this section we introduce the basic ideas of autoencoders, and then show our results.

\subsection{Autoencoders}

Autoencoder is a type of neural network that learns to reconstruct input data in an unsupervised manner \citep{Rumelhart1986LearningRB, 2022arXiv220103898M}. It comprises two components: an encoder and a decoder. The encoder is a feed-forward neural network that compresses the input data into a lower-dimensional space, also known as latent space. The decoder performs the inverse process, transforming the data from the latent space back to the input dimension. The bottleneck layer, located at the last layer of the encoder, has some dimensionality constrains that guarantee that it has fewer neurons than the previous layers, to ensure that the autoencoder does not merely copy the input data.

The power of autoencoders lies in their ability to capture the primary features of the input data, commonly known as latent features. This is akin to Principal Component Analysis (PCA), but while PCA is a linear method, autoencoders are not restricted to linear relationships on the data and can capture more complex features. Autoencoders have a wide range of applications, including image denoising, feature extraction, and anomaly detection. They are particularly effective in certain image anomaly detection problems, due to their ability to work in an unsupervised way and therefore not requiring previous knowledge about the anomalies. Based on the assumption that the amount of anomalous data is very small compared to the size of the entire dataset, autoencoders automatically learn the features that describe ``normal'' data.

\subsection{Training data and network}
Our dataset consists of 10000 source stamps from DECaLS DR3, each of size 64x64. We include sources from the g, r, and z bands.

\begin{figure}
    \includegraphics[width=\columnwidth]{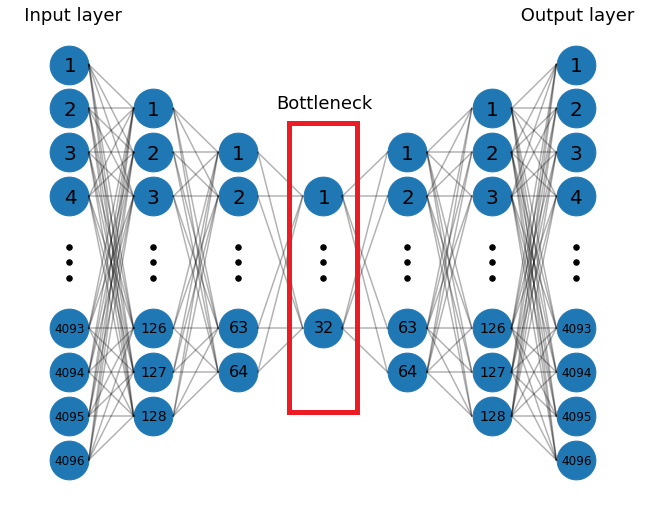}
    \caption{Architecture of our autoencoder. The encoder has three layers, with 128, 64, and 32 nodes respectively. The decoder has three layers with 32, 64, and 128 nodes, respectively.}
	\label{fig:autoencoder_arquitechture}
\end{figure}

The input of our neural network are the flattened source stamps, with size 4096 ($64*64$). The encoder is composed by three dense layers with 128, 64, and 32 nodes respectively. The decoder is also composed by three dense layers with 32, 64, and 128 nodes, respectively. The output of the decoder has equal size to the input data. Fig.~\ref{fig:autoencoder_arquitechture} presents a visual representation of our Autoencoder. Although this is a very simple model, it suffices for our purpose since the source stamps do not exhibit complex features and we aim to avoid overfitting.

We divide our dataset into training ($\sim 80\%$) and validation ($\sim 20\%$) sets and train our network for 500 epochs, reaching a stable validation loss. We select the model at the epoch with lowest validation loss as our final model to make predictions.

\subsection{Results} \label{section:results_autoencoder}
For each exposure in DECaLS DR3, we select all its sources and we feed them one by one to our trained model. The output of the model is the reconstructed version of the input source stamp. We calculate the mean squared error (MSE) between the original stamp and its reconstructed counterpart as:
\begin{equation}
    \text{MSE} = \frac{1}{N^2} \sum_{i=1}^{N} \sum_{j=1}^{N} \left(\hat{x}_{ij}-x_{ij}\right)^2,
\end{equation}
where $x_{ij}$ is the (i, j) pixel of the original source stamp and $\hat{x}_{ij}$ is the (i, j) pixel of the reconstructed source stamp. $N=64$ is the stamp size.

We finally calculate the mean MSE of all the sources within each exposure and use that value as our source anomaly indicator. A higher mean MSE indicates a higher likelihood of that exposure being anomalous.

\begin{figure}
    \begin{center}
    \includegraphics[width=0.49\textwidth]{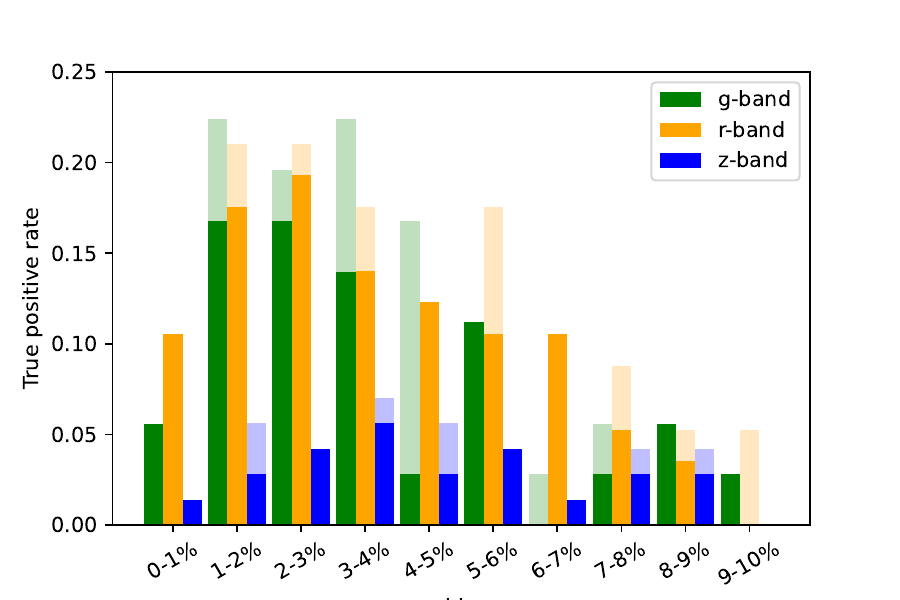}
	\caption{True positive rate for the 10\% of exposures with highest likelihood of presenting source anomalies according to our deep learning method. The true positive rate is based on human inspection. We divide the exposures into 10 bins of equal size, for each band (g, r, z). Dark colors include only prominent anomalies, whereas light colors also include dubious anomalies.}
	\label{fig:results_sources_cnn}
     \end{center}
\end{figure}

Fig.~\ref{fig:results_sources_cnn} shows our results. We plot our results as in fig.~\ref{fig:results_sources} and fig.~\ref{fig:results_back}. Dark colors include only prominent anomalies whereas light colors also include dubious ones. Our model shows a good performance in detecting source anomalies, particularly the prominent ones. This is expected since prominent anomalies strongly distort the sources, and these look clearly different from all sources within normal images. However, dubious anomalies induce small elongations on the sources. These small elongations only become problematic when we look at the entire exposure, not to a single source. Single sources have intrinsic shapes of the same order as the elongations, and therefore single sources elongated by the anomaly simply look like normal sources, and their reconstruction is very accurate.

To conclude, although the autoencoder shows a decent performance in detecting source anomalies, it is not as effective as the Entropy Method. The autoencoder fails to detect many anomalous exposures that were identified by the Entropy Method and it shows a higher false positive rate.

\section{Conclusions} \label{section:Conclusion}

The detection of anomalies is a crucial aspect of astronomical research, as it enables scientists to identify and understand unusual phenomena that deviate from expected patterns. These anomalies represent observational artifacts that need to be carefully distinguished from genuine astronomical signals.

In this study, we have presented two methods for detecting anomalous images in astronomical datasets. We tested our methods on the i-band of CFHTLenS and the g, r, and z bands of DECaLS DR3. Our first method, named as Entropy Method, is based on the fact that both source and background anomalies introduce systematic anisotropies on the CCD images. Leveraging this insight, we study the statistical orientations of sources to detect source anomalies and the statistical orientation of the gradients at background points to identify background anomalies. The Entropy Method has demonstrated excellent results in detecting anomalies in both CFHTLenS and DECaLS DR3. In CFHTLenS, the Entropy Method outperforms human inspection by detecting 12 of the 13 anomalous exposures previously identified by human inspection and detecting ten new ones. In DECaLS DR3, the Entropy Method has detected a substantial number of source and background anomalies while maintaining a low false positive rate.
Our second method involved training an autoencoder to reconstruct source stamps, and we used the quality of these reconstructions as an indicator of the likelihood of exposures containing source anomalies. After evaluating this method on DECaLS DR3, we found that although the autoencoder performs relatively well, the performance of the Entropy Method is significantly better. The autoencoder fails to detect several anomalous exposures identified by the Entropy Method and exhibits a higher false positive rate.

Autoencoders are unsupervised, meaning that they do not use any information about the anomalies for training. We chose an unsupervised method to ensure fair comparisons with the Entropy Method. However, utilizing the anomalous exposures found by the Entropy Method to train a neural network in a supervised way could potentially increase the performance of anomaly detection. This would lead to a more general method that could combine the anomalous exposures from different datasets to perform a more generalized anomaly detection method, applicable to new datasets. Once new anomalous exposures are found, the neural network could be fine-tuned. This approach is left for a future work.

Overall, we believe that our research serves as a stepping stone for future advancements in anomaly detection techniques in astronomy. Our work lays the foundation for more sophisticated and efficient methods, potentially combining traditional statistical methods and machine learning methods to further enhance anomaly detection capabilities. As the field of astronomy continues to evolve with technological advancements and increasingly vast datasets, our methods are poised to play valuable roles in uncovering the secrets of the cosmos.

\section{acknowledgements}
CFHTLenS is based on observations obtained with MegaPrime/MegaCam, a joint project of CFHT and CEA/IRFU, at the Canada-France-Hawaii Telescope (CFHT) which is operated by the National Research Council (NRC) of Canada, the Institut National des Science de l'Univers of the Centre National de la Recherche Scientifique (CNRS) of France, and the University of Hawaii. This work is based in part on data products produced at Terapix available at the Canadian Astronomy Data Centre as part of the Canada-France-Hawaii Telescope Legacy Survey, a collaborative project of NRC and CNRS.

The Legacy Imaging Surveys of the DESI footprint is supported by the Director, Office of Science, Office of High Energy Physics of the U.S. Department of Energy under Contract No. DE-AC02-05CH1123, by the National Energy Research Scientific Computing Center, a DOE Office of Science User Facility under the same contract; and by the U.S. National Science Foundation, Division of Astronomical Sciences under Contract No. AST-0950945 to NOAO. The Photometric Redshifts for the Legacy Surveys (PRLS) catalog used in this paper was produced thanks to funding from the U.S. Department of Energy Office of Science, Office of High Energy Physics via grant DE-SC0007914.

JZ is supported by the National Key Basic Research and Development Program of China (No.2018YFA0404504), and the NSFC grants (11621303, 11890691, 12073017), the science research grants from China Manned Space Project (No. CMS-CSST-2021-A01). XDL is supported by the China Manned Space Project with No. CMS-CSST-2021-A03 and No. CMS-CSST-2021-B01.

The computations in this paper were run on the $\pi$ 2.0 cluster supported by the Center of High Performance Computing at Shanghai Jiaotong University, and the Gravity supercomputer of the Astronomy Department, Shanghai Jiaotong University.

\bibliography{main}{}
\bibliographystyle{aasjournal}

\clearpage

\end{document}